\documentclass[10pt,conference]{IEEEtran}
\usepackage{graphicx,subfigure}
\usepackage{amsfonts,amsmath,amssymb}
\usepackage{epic,eepic,eepicemu}
\usepackage{epsf}
\usepackage{epsfig}
\usepackage{fancyheadings}
\usepackage{graphics}
\usepackage{psfrag}

\makeatletter
\def\@cite#1#2{[\if@tempswa #2 \fi #1]}
\makeatother

\newcommand{\myparagraph}[1]{{\bf{#1}}}

\newcommand{\widgraph}[2]{\includegraphics[keepaspectratio,width=#1]{#2}}

\newcommand{\beq}{\begin{equation}}
\newcommand{\eeq}{\end{equation}}
\newcommand{\bea}{\begin{eqnarray}}
\newcommand{\eea}{\end{eqnarray}}

\def\qed{\quad \vrule height6.5pt width6pt depth0pt} %Nifty end proof sign
\def\qed{\quad \vrule height6.5pt width6pt depth0pt} %Nifty end proof sign
\newtheorem{atheorem}{Theorem}
\newtheorem{lemma}{Lemma}
\newtheorem{definition}{Definition}
\newtheorem{proposition}{Proposition}
\newtheorem{corollary}{Corollary}

\long\def\symbolfootnote[#1]#2{\begingroup
\def\thefootnote{\fnsymbol{footnote}}\footnote[#1]{#2}\endgroup}

\newcommand{\const}{\ensuremath{C}}
\newcommand{\nbit}{\ensuremath{n}}
\newcommand{\mcheck}{\ensuremath{m}}
\newcommand{\Neigh}{\ensuremath{N}}

\newcommand{\actset}{\ensuremath{\mathbb{A}}}
\newcommand{\cw}{\ensuremath{x^{\operatorname{cw}}}}
\newcommand{\pcw}{\ensuremath{x^{\operatorname{pc}}}}
\newcommand{\rate}{\ensuremath{R}}
\newcommand{\cdeg}{\ensuremath{d_c}}
\newcommand{\vdeg}{\ensuremath{d_v}}
\newcommand{\vertex}{\ensuremath{V}}

\newcommand{\mybeginproof}{\noindent \emph{Proof: $\;$}}
\newcommand{\myendproof}{\hfill \qed}

\newcommand{\vertfrac}{\ensuremath{\vertex_{\operatorname{frac}}}}
\newcommand{\checkfrac}{\ensuremath{C_{\operatorname{frac}}}}

\newcommand{\Forbid}{\ensuremath{\mathbb{F}}}
\newcommand{\Boxin}{\ensuremath{\mathbb{B}}}

\newcommand{\checknum}{\ensuremath{m}}
\newcommand{\bitnum}{\ensuremath{n}}
\newcommand{\plaincw}{\ensuremath{x}}
\newcommand{\CheckMat}{\ensuremath{H}}
\newcommand{\Code}{\ensuremath{\mathbb{C}}}

\newcommand{\RelPoly}{\ensuremath{\mathcal{P}}}

\long\def\comment#1{}

\begin{document}

\title{\huge{Guessing Facets: Polytope Structure and Improved LP
Decoder}}


\author{Alexandros G. Dimakis$^1$ and Martin J. Wainwright$^{1,2}$ \\
$^1$ Department of Electrical Engineering and Computer Science \\ $^2$
Department of Statistics \\ University of California, Berkeley\\
\texttt{\small $\{$adim,wainwrig$\}$@eecs.berkeley.edu} \\ }

% avoiding spaces at the end of the author lines is not a problem with
% conference papers because we don't use \thanks or \IEEEmembership

\maketitle

\begin{center}
%% PREPRINT LABEL:
\vspace*{-2.5in}
\vbox to 2.5in{\footnotesize {\tt
 \begin{tabular}[t]{c}
    Appeared in: \\
    International Symposium on Information Theory \\
    Seattle, WA;  July 2006
  \end{tabular} \vfil}}
\end{center}

\begin{abstract}

A new approach for decoding binary linear codes by solving a linear
program (LP) over a relaxed codeword polytope was recently proposed by
Feldman et al.  In this paper we investigate the structure of the
polytope used in the LP relaxation decoding. 
%This relaxed polytope has
%both integral vertices which correspond to codewords and non-integral
%vertices called fractional pseudocodewords.  
We begin by showing that for expander codes, every fractional
pseudocodeword always has at least a constant fraction of non-integral
bits.  We then prove that for expander codes, the active set of any
fractional pseudocodeword is smaller by a constant fraction than the
active set of any codeword.  We exploit this fact to devise a decoding
algorithm that provably outperforms the LP decoder for finite
blocklengths.  It proceeds by guessing facets of the polytope, and
resolving the linear program on these facets. While the LP decoder
succeeds only if the ML codeword has the highest likelihood over all
pseudocodewords, we prove that for expander codes the proposed
algorithm succeeds even with a constant number of pseudocodewords of
higher likelihood.  Moreover, the complexity of the proposed algorithm
is only a constant factor larger than that of the LP decoder.
%We further give some results about the structure of the
%relaxed polytope and bounds on the number of pseudocodewords.

\end{abstract}

\section{Introduction}

Low-density parity check (LDPC) codes are a class of graphical codes,
originally introduced by Gallager~\cite{Gallager63}, that are known to
approach capacity as the blocklength increases, even when decoded with
the sub-optimal sum-product algorithm.  The standard techniques for
analyzing the sum-product algorithm, including density
evolution~\cite{Richardson01a} and EXIT charts~\cite{Ashikhmin04}, are
asymptotic in nature.  Many applications, however, require the use of
intermediate blocklengths, in which regime asymptotic analysis methods
are not suitable for explaining or predicting the behavior of the
decoding algorithms.  Recently, Feldman et al.~\cite{Feldman05}
introduced the LP decoding method, which is based on solving a
linear-programming relaxation of the ML decoder method.  While LP
decoding performance is not better to message-passing decoders, a
possible advantage is its relative amenability to finite-length
analysis. \\

\noindent \myparagraph{Previous work:} The LP decoding idea was
introduced by Feldman et al.~\cite{Feldman03,Feldman05}.  There are
various theoretical connections between LP decoding and
message-passing~\cite{FelKarWai02,KoeVon03,WaiJaaWil05b}.  For the
binary symmetric channel, it can be shown~\cite{Felplus03} that LP
decoding can correct a linear fraction of errors for suitable expander
codes.  Vontobel and Koetter~\cite{kv_bethe,KoeVon03} established
bounds on the pseudo-weight for Gaussian channels, showing that it
grows only sublinearly for regular codes.  Feldman and
Stein~\cite{FeldmanStein05} proved that LP decoding can achieve
capacity when applied to generalized expander constructions.

\noindent \myparagraph{Our contributions:} The LP decoder operates by
solving a linear program over a polytope $\mathcal{P}$ which
constitutes a relaxation of the original combinatorial codeword
space. The polytope $\mathcal{P}$ has two types of vertices:
\emph{integral vertices} with $0-1$ components corresponding to
codewords, and \emph{fractional vertices} that correspond to
pseudocodewords.  This paper begins by studying the geometric
properties of the relaxed polytope.  In particular, we prove that for
suitable classes of expander codes, the relaxed polytope $\mathcal{P}$
has the property that more facets are adjacent to integral points
relative to fractional ones.  Motivated by this geometric intuition,
we propose an improved LP decoding algorithm that eliminates
fractional pseudocodewords by guessing facets of $\mathcal{P}$, and
then decodes by re-solving the optimization problem on these
facets. We prove some theoretical results on the performance of this
facet-guessing decoder.  Our experimental results show significant
performance improvements, particularly at high SNR, for small and
moderate blocklengths.

\section{Background}

Consider a binary linear code with $\bitnum$ bits and $\checknum$
checks, and let $\rate = 1 - \frac{\checknum}{\bitnum}$.  It can
be specified by a parity check matrix $\CheckMat \in
\{0,1\}^{\checknum \times \bitnum}$: in particular, the code
$\Code$ consists of all vectors $\plaincw \in \{0,1\}^\bitnum$
that satisfy $\CheckMat \plaincw = 0$, where multiplication and
addition are performed over $GF(2)$. \\

\noindent \myparagraph{Maximum likelihood decoding as a linear
program:} The codeword polytope of a code is the convex hull of all
its codewords. Maximum likelihood (ML) decoding can be written as a
linear program involving the codeword polytope but unfortunately there
are no known ways for describing the codeword polytope efficiently.
In fact, the existence of a polynomial-time separation oracle for the
codeword polytope of a general linear code is very unlikely since ML
decoding for arbitrary linear codes is NP-hard~\cite{Berlekamp78}. \\

\noindent \myparagraph{Relaxed polytope and LP decoding:} The relaxed
polytope $\mathcal{P}$ is an approximation to the codeword polytope
that can be described by a linear number of inequalities for LDPC
codes.  For each check, the corresponding local codeword polytope
(LCP) is the convex hull of the bit sequences that satisfy the check
(local codewords). For checks of constant bounded degree, the LCP can
be described by a constant number of inequalities. The relaxed
polytope $\mathcal{P}$ is obtained by looking at each check
independently, and taking the intersection of all the local codeword
polytopes.

More specifically, for every check we can find the bit sequences that
violate it (local forbidden sequences) and make sure we are
sufficiently far away from them. So for every check $j$ connected to
variables $N(j)$ find all the possible forbidden sequences $S$ and
make sure that their $\ell_1$ distance is at least one---viz.
$\sum_{N(j)\backslash S} f_i + \sum _{i\in S} (1-f_i) \geq 1$.  It can
be shown that by picking the $\ell_1$ distance to be one we are not
excluding any legal codewords from our relaxed polytope.  We will call
these constraints \emph{forbidden set inequalities}.  We also need to
add $2 n$ inequalities $0\leq f_i \leq 1$, denoted \emph{box
inequality constraints}, which ensure that $f$ remains inside the unit
hypercube.  It can be shown that for every check, the set of its
forbidden inequalities along the box inequalities for the associated
variables, describe the LCP of the check.
%Sometimes (for
%example in checks of degree 3) the box inequalities are redundant, but
%this is not always the case.  
The relaxed polytope is defined as the intersection of all the LCPs
(i.e., the constraints consist of all forbidden set inequalities
$\Forbid$ along with the box inequalities).

Notice that for every check with degree $d_c$ there is an exponential
number of sequences $2^{d_c-1}$ of local forbidden sequences and
therefore the total number of forbidden sequences is $2^{d_c -1
}m$. For low-density parity-check codes, $d_c$ is either fixed (for
regular) or small with high probability (for irregular) so the number
of local forbidden sequences is linear in blocklength.  Therefore the
relaxed polytope can be described by a linear number of inequalities.

%(\textbf{however yannakakis construction works for general
%linear codes..}).

%Figure 1 illustrates the codeword polytope for one
%check $x_1 \oplus x_2 \oplus x_3=0$.
%The codewords of this trivial
%code are illustrated in red. In this example, the codeword
%polytope (the convex hull of the codewords) and the relaxed
%polytope (the set defined by inequalities \ref{inequalities}) are
%identical.

Finally, it can be shown that if the LDPC graph had no cycles, the
local forbidden sequences would identify all the possible
non-codewords and the relaxation would be exact.  However if the graph
has cycles, there exist vertices with non $\{0,1\}$ coordinates that
satisfy all the local constraints individually and yet are not
codewords nor linear combinations of codewords.  These sequences are
called (fractional) pseudocodewords. To simplify the presentation, we
will call all the vertices of the relaxed polytope pseudocodewords (so
codewords are also pseudocodewords) and fractional pseudocodewords
will be the vertices of the relaxed polytope which happen to have at
least one fractional coordinate. One question relates to the number of
fractional coordinates (fractional support) that a pseudocodeword can
have. While codes can be constructed that have an arbitrarily small
fractional support, we show that for expander codes, the fractional
support has size at least linear in blocklength. Using this result, we
show that for expander codes, the active set of any fractional
pseudocodeword (i.e., the number of inequalities that are active at
the vertex) is smaller than the active set size of any codeword by at
least a linear fraction (in blocklength). These results naturally lead
to a randomized algorithm for improving the performance of the
LP-decoder by guessing facets of the relaxed polytope and resolving
the optimization problem.

\section{Structure of the relaxed polytope}

\begin{definition}  A $(d_c, d_v)$-regular bipartite graph
is an $(\alpha, \delta)$ expander if, for all subsets $|S| \leq \alpha
\nbit$, there holds $|\Neigh(S)| \geq \delta d_v |S|$.
\end{definition}

\subsection{Fractional support of pseudocodewords}

A quantity of interest is the fractional support of a pseudocodeword,
defined as follows.
\begin{definition}
The fractional support of a pseudocodeword $\pcw$ is the subset
$\vertfrac(\pcw) \subseteq \vertex$ of bits indices in which $\pcw$
has fractional elements.  Similarly, the subset of checks that are
adjacent to fractional elements of $\pcw$ is denoted by
$\checkfrac(\pcw)$.
\end{definition}

The following result dictates that all pseudocodewords in an expander
code have substantial fractional supports:
\begin{proposition}
\label{PropFracSupport} Given an $(\alpha, \delta)$-expander code
with \mbox{$\delta > \frac{1}{2}$,} any pseudocodeword has fractional
support that grows linearly in blocklength:
\begin{eqnarray*}
|\vertfrac(\pcw)| \; \geq \; \alpha \nbit, \quad \mbox{and} \quad
|\checkfrac(\pcw)| \; \geq \; \delta \vdeg \alpha \nbit.
\end{eqnarray*}
\end{proposition}
\mybeginproof The proof is based on a series of lemmas:
\begin{lemma}[Unique neighbor property~\cite{ZipSpi96}]
\label{LemUniqueNeigh}
Given an $(\alpha, \delta)$ expander with $\delta > \frac{1}{2}$, any
subset $S \subseteq V$ of size at most $\alpha n$ satisfies the unique
neighbor property, i.e there exists $y \in C$ such that $|\Neigh(y)
\cap S|=1$.
\end{lemma}
\mybeginproof Proceed via proof by contradiction: suppose that every
$y \in \Neigh(S)$ has two or more neighbors in $S$.  Then the total
number of edges arriving at $\Neigh(S)$ from $S$ is at least $2
|\Neigh(S)| > 2 \delta \vdeg |S| > \vdeg |S|$. But the total number of
edges leaving $S$ has to be exactly $\vdeg |S|$, which yields a
contradiction. \myendproof

\begin{lemma}
\label{LemSingFrac}
In any pseudocodeword $\pcw$, no check is adjacent to only one
fractional variable node.
\end{lemma}
\mybeginproof Suppose that there exists a check adjacent to only one
fractional bit: then the associated local pseudocodeword is in the
local codeword polytope (LCP) for this check and therefore can be
written as a linear combination of two or more
codewords~\cite{Ziegler}.  But these local codewords would have to
differ in only one bit, which is not possible for a parity check.
\myendproof

We can now prove the main claim.  Consider any set $S$ of fractional
bits of size $|S| \leq \alpha \nbit$.  Using the expansion and
Lemma~\ref{LemUniqueNeigh}, the set $\Neigh(S)$ must contain at least
one check adjacent to only one bit in $S$.  By
Lemma~\ref{LemSingFrac}, this check must be adjacent to at least one
additional fractional bit.  We then add this bit to $S$, and repeat
the above argument until $|S| > \alpha \nbit$, to conclude that
$|\vertfrac(\pcw)| > \alpha \nbit$.  Finally, the bound on
$|\checkfrac(\pcw)|$ follows by applying the expansion property to a
subset of fractional bits of size less than or equal to $\alpha
\nbit$.
\myendproof

\subsection{Sizes of active sets}

For a vertex $v$ of a polytope, its active set $\actset(v)$ is the
set of linear inequalities that are satisfied with equality on
$v$. Geometrically, this corresponds to the set of facets of the
polytope that contain the vertex $v$. We want to determine the
size of active sets for codewords and pseudocodewords. The key
property we want to prove is that for expander codes, codewords
have active sets which are larger by at least a constant factor.

\begin{atheorem}
\label{ThmPCWSize} For any $(\vdeg, \cdeg)$ code with $\rate \in
(0,1)$, the active set of any codeword $\cw$ has
\begin{eqnarray}
\label{EqnActsetCW} |\actset(\cw)| & = \gamma_{cw} \nbit.
\end{eqnarray}
elements.  For an $(\alpha, \delta)$-expander code with $\delta >
\frac{1}{2}$, the active set of any fractional pseudocodeword $\pcw$
is smaller than the active set of any codeword by a linear
fraction---in particular,
\begin{eqnarray}
\label{EqnActsetPCW} |\actset(\pcw)| & \leq & \nbit \gamma_{pc}
 % \nonumber \\
%%
%& = & |\actset(\cw)| - \nbit \biggr \{ \alpha \big[ \vdeg \delta
%  \big(\cdeg-2 \big) +1 \big] \biggr\}.
\end{eqnarray}
where the constants are \mbox{$\gamma_{cw}=\big[ (1-\rate )\cdeg +1
\big]$} and \mbox{$\gamma_{pc}= \Big[ \big( 1- \rate -\delta \vdeg
\alpha \big) \cdeg + 2 \delta \vdeg \alpha + (1 - \alpha)\Big]$.}
(Note that $\gamma_{pc} < \gamma_{cw}$.)
\end{atheorem}
\mybeginproof
We begin by proving equation~\eqref{EqnActsetCW}.  By the
code-symmetry of the relaxed polytope~\cite{Feldman05}, every codeword
has the same number of active inequalities, so it suffices to restrict
our attention to the all-zeroes codeword.  The check inequalities
active at the all-zeros codeword are in one-to-one correspondence with
those forbidden sequences at Hamming distance $1$.  Note that there
are $\cdeg$ such forbidden sequences, so that the total number of
constraints active at the all-zeroes codeword is simply
$|\actset(\cw)| = \mcheck \cdeg + \nbit \; = \; \nbit \, \big[
(1-\rate) \cdeg + 1 \big]$ as claimed.

We now turn to the proof of the bound~\eqref{EqnActsetPCW} on the size
of the fractional pseudocodeword active set.  Recall that the relaxed
polytope consists of two types of inequalities: \emph{forbidden set
constraints} (denoted $\Forbid$) associated with the checks, and the
\emph{box inequality constraints} $0 \leq x_i \leq 1$ (denoted
$\Boxin$) associated with the bits.  The first ingredient in our
argument is the fact (see Proposition~\ref{PropFracSupport}) that for
an $(\alpha, \delta)$-expander, the fractional support
$\vertfrac(\pcw)$ is large, so that a constant fraction of the box
inequalities will not be active.

Our second requirement is a bound on the number of forbidden set
inequalities that can be active at a pseudocodeword.  We establish a
rough bound for this quantity using the following lemma:
\begin{lemma}
\label{LemSurf1}
Suppose that $z$ belongs to a polytope and is not a vertex.  Then
there always exist at least two vertices $x, y$ such that $\actset(z)
\subseteq \actset(x) \cap \actset(y)$.
\end{lemma}
\mybeginproof Since $z$ belongs to the polytope but is not a vertex,
it must either belong to the interior, or lie on a face with dimension
at least one.  If it lies in the interior, then \mbox{$\actset(z)=
\emptyset$,} and the claim follows immediately.  Otherwise, $z$
must belong to a face $F$ with $\dim(F) \geq 1$.  Then $F$ must
contain~\cite{Ziegler} at least $\dim(F) + 1 = 2$ vertices, say $x$
and $y$.  Consequently, since $x, y$ and $z$ all belong to $F$ and $z$
is not a vertex, we must have $\actset(z) \subseteq \actset(y)$ and
$\actset(z) \subseteq \actset(x)$, which yields the claim.
\myendproof

Given a check $c$ and codeword $\cw$, let $\Pi_c(\cw)$ denote the
restriction of $\cw$ to bits in the neighborhood of $c$ (i.e., a
\emph{local codeword} for the check $c$).  With this notation, we
have:
\begin{lemma}
\label{LemSurf2}
For any two local codewords $\Pi_c(\cw_1)$ and $\Pi_c(\cw_2)$ of a
 check $c$, the following inequality holds
\begin{equation}
|\actset(\Pi_c(\cw_1)) \cap \actset(\Pi_c(\cw_2))| \leq 2.
\end{equation}
\end{lemma}
\mybeginproof
The intersection $\actset(\Pi_c(\cw_1)) \cap \actset(\Pi_c(\cw_2))$ is
given by the forbidden sequences that have Hamming distance $1$ from
$\Pi_c(\cw_i), i = 1,2$ (i.e., forbidden sequences $f$ such that $d(f,
\Pi_c(\cw_i)) = 1$ for $i=1,2$).  Thus, if such an $f$ exists, then by
the triangle inequality for Hamming distance, we have
\begin{equation}
2 = d(f,\Pi_c(\cw_1)) + d(f, \Pi_c(\cw_2))) \geq d(\Pi_c(\cw_1),
\Pi_c(\cw_2)),
\end{equation}
But $d(\Pi_c(\cw_1), \Pi_c(\cw_2)) \geq 2$ for any two local
codewords, so that we must have $d(\Pi_c(\cw_1), \Pi_c(\cw_2)) = 2$.
Consequently, we are looking for all the forbidden (odd) sequences of
length $\cdeg$ that differ in one bit from two local codewords that
are different in two places.  Clearly there are only two such
forbidden sequences, so that the claim follows.
\myendproof \\

We can now establish a bound on the size of the active sets of
pseudocodewords for $(\alpha, \delta)$-expanders:
\begin{lemma}
\label{LemUpperBound}
For every pseudocodeword $\pcw$, the size of the active set 
$|\actset(\pcw)|$ is upper
bounded by
\begin{equation}
\label{EqnActiveSet}
(\mcheck -|\checkfrac(\pcw)|) \cdeg + 2
|\checkfrac(\pcw)| 
+ \nbit - |\vertfrac(\pcw)|.
\end{equation}
\end{lemma}
\mybeginproof The proof is based on the decomposition:
\begin{eqnarray*}
|\actset(\pcw)| & = & |\actset(\pcw) \cap \Forbid| + \actset(\pcw)
|\cap \Boxin|.
\end{eqnarray*}
The cardinality $|\actset(\pcw) \cap \Boxin|$ is equal to the number
of integral bits in the pseudocodeword, given by $\nbit -
|\vertfrac(\pcw)|$.

We now turn to upper bounding the cardinality $|\actset(\pcw) \cap
\Forbid|$.  Consider the $\mcheck - |\checkfrac(\pcw)|$ checks that
are adjacent to only integral bits of $\pcw$.  For each such check,
exactly $\cdeg$ forbidden set constraints are active, thereby
contributing a total of $\cdeg \big[\mcheck - |\checkfrac(\pcw)|\big]$
active constraints.  Now consider one of the remaining
$|\checkfrac(\pcw)|$ fractional checks, say $c$.  Consider the
restriction $\Pi_c(\pcw)$ of the pseudocodeword $\pcw$ to the check
neighborhood of $c$.  Since $\Pi_c(\pcw)$ contains fractional
elements, it is not a vertex of the local codeword polytope associated
with $c$.  Therefore, by combining Lemmas~\ref{LemSurf1}
and~\ref{LemSurf2}, we conclude that $|\actset(\Pi_c(\pcw))| \leq 2$.
Overall, we conclude that the upper bound~\eqref{EqnActiveSet} holds.
\myendproof \\

Using Lemma~\ref{LemUpperBound} and Proposition~\ref{PropFracSupport},
we can now complete the proof of Theorem~\ref{ThmPCWSize}.  In
particular, we re-write the RHS of the bound~\eqref{EqnActiveSet} as
%\begin{equation*}
\mbox{$(1-\rate)\cdeg \, \nbit - (\cdeg - 2) |\checkfrac(\pcw)| +
  \nbit - |\vertfrac(\pcw)|$.}
%\end{equation*}
From Proposition~\ref{PropFracSupport}, we have
$|\checkfrac(\pcw)| \geq \vdeg \delta \alpha \nbit$ and
$|\vertfrac(\pcw)| > \alpha \nbit$, from which the
bound~\eqref{EqnActsetPCW} follows.

\myendproof
%

%
%\comment{ A convenient way to represent the combinatorial
%structure\footnote{not sure if it captures the complete combinatorial
%structure, Ziegler (chap 0) says so but it is for an example of a
%cyclic polytope} of the relaxed polytope $\mathcal{F}$ is to use the
%vertex-facet bipartite graph\cite{Ziegler} $\mathcal{B}$ that has the
%set of pseudocodewords (vertices of $\mathcal{F}$) $V$ on one class
%and the set of linear inequalities or facets $A$ on the other. We
%connect a pseudocodeword to a facet if it is contained in that facet
%in $\mathcal{F}$. For an illustration see figure \ref{VF_graph}. The
%degree of vertices in $\mathcal{B}$ that correspond to codewords is
%$|A_c|$ and the degree for a fractional pseudocodeword $p$ is
%$|A_p|$. Notice that the the number of pseudocodewords $|V|$ is
%exponential (as we demonstrate in subsequent section) while the number
%of facets $|A|$ is only linear in $n$.
%}

\begin{figure}
\begin{center}
\psfrag{#ac#}{$\actset(\cw)$} \psfrag{#ap#}{$\actset(\pcw)$}
\psfrag{#vf#}{$\vertex(F)$} \psfrag{#vert#}{vertices}
\psfrag{#fac#}{facets}
\widgraph{.20\textwidth}{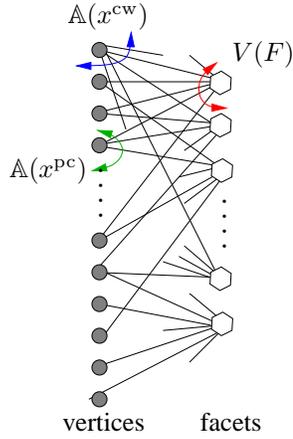}
\end{center}
\caption{Vertex-facet diagram of the relaxed polytope.  Circles on the
left-hand side correspond to vertices (codewords $\cw$ and fractional
pseudocodewords $\pcw$) of the relaxed polytope; hexagons on the
right-hand side correspond to facets (hyperplane inequalities)
defining the relaxed polytope. }
\label{FigVertexFacet}
\end{figure}

\section{Improved LP decoding}

Various improvements to the standard sum-product decoding algorithm
have been suggested in past work~\cite[e.g.,]{Fossorier01,PishroFekri}.
Based on the structural results that we have obtained, we now describe
some improved decoding algorithms for which some finite-length
analysis is possible.  We begin with some simple observations: (i) ML
decoding corresponds to finding the vertex in the relaxed polytope
that has the highest likelihood and integral coordinates; and (ii)
Standard LP decoding succeeds if and only if the ML codeword has the
highest likelihood over all pseudocodewords.

These observations highlight the distinction between LP decoding and
ML decoding.  An LP solver, given the (polynomially many) facets of
the relaxed polytope, determines the vertex with the highest
likelihood without having to go through all the exponentially many
vertices of $V$. In contrast, the ML decoder can go down this list,
and determine the first vertex which has integral coordinates.  This
motivates facet-guessing: suppose that there exists only one
fractional pseudocodeword $\pcw_1$ that has higher likelihood than the
ML codeword $\cw$.  The LP decoder will output the pseudocodeword
$\pcw_1$, resulting in a decoding error.  However, now suppose that
there exists a facet $F_1 \in \actset$ such that $\cw \in F_1$ but
$\pcw \notin F_1$.  Consider the reduced polytope $\mathcal{P'}$
created by restricting the relaxed polytope $\mathcal{P}$ to the facet
$F_1$ (i.e., $\mathcal{P'} = \RelPoly \cap F_1$).  This new polytope
will have a vertex-facet graph $\mathcal{B'}$ with vertices $V'=
N(F_1)$ i.e. all the vertices that are contained in $F_1$. The
likelihoods will be the same, but $p_1$ will not belong in
$\mathcal{P'}$ and therefore we can use an LP solver to determine the
vertex with the highest likelihood in $\mathcal{P'}$ which will be
$c$. Therefore if we could guess the right facet $F_1$ we can
determine the ML codeword for this case.  Based on this intuition, we
introduce two postprocessing algorithms for improving LP decoding.

{\bf{Facet Guessing Algorithm}}
\begin{enumerate}
\item Run LP decoding: if outputs an integral codeword, terminate.
Otherwise go to Step 2.
\item Take as input:
\begin{itemize}
\item fractional pseudocodeword $\pcw$ from the LP decoder
\item likelihood vector $\gamma$.
\end{itemize}

\item Given a natural number$ N\geq 1$, repeat for $i=1, \ldots N$:
\begin{enumerate}
\item[(a)] Select a facet $F_i \in (\actset \setminus
\actset_{\pcw})$, form the reduced polytope a new polytope
$\mathcal{P'} = \RelPoly \cap F_i$.

\item[(b)] Solve the linear program with objective vector $\gamma$ in
$\mathcal{P'}$, and save the optimal vertex $z_i$.
\end{enumerate}

\item From the list of optimal LP solutions $\{z_1, \ldots, z_N \}$,
output the integral codeword with highest likelihood.
\end{enumerate}

%%%%%%%%%%%%%%%%%%%%%%%%%%%%

\myparagraph{Remarks:} (a) There are two variations of facet guessing:
exhaustive facet guessing (EFG) tries all possible facets (i.e., $N =
|(\actset \setminus \actset_{\pcw})|$), while randomized facet
guessing (RFG) randomly samples from $(\actset \setminus
\actset_{p_i})$ a constant number of times (e.g., $N = 20$). (b) Note
that the EFG algorithm has polynomial-time complexity.  Since
$|\actset \setminus \actset_{\pcw}|=O(n)$ this requires only a linear
number of calls to an LP solver. On the other hand, the RFG algorithm
requires a constant number of calls to an LP solver and therefore has
the same complexity order as LP decoding.  We now provide a
characterization of when the EFG algorithm fails:
\begin{lemma}
The exhaustive facet-guessing algorithm fails to find the ML codeword
$c$ $\iff$ every facet $F \in \actset_c$ contains a fractional
pseudocodeword with likelihood greater than $c$.
\label{exhaustive}
\end{lemma}
\mybeginproof Denote the set of fractional pseudocodewords with
likelihood higher than $c$ by $\hat{p}$. Assume there exists a facet
$F_i$ such that $c \in F_i$ and $\forall p \in \hat{p}$, $p \notin
F_i$. Then the algorithm will at some point select $F_i$ and the LP
solver will output the vertex in $\mathcal{P'}$ with the highest
likelihood which will be $c$ since nothing from $\hat{p}$ can belong
in $\mathcal{P'}$.  Therefore $c$ will be in the list of LP
solutions. Also, since $c$ is the ML codeword, there can be no other
integral codeword with higher likelihood in the list, and therefore
the algorithm will output $c$.
\myendproof

\begin{figure*}
\begin{center}
\begin{tabular}{cc}
\widgraph{.38\textwidth}{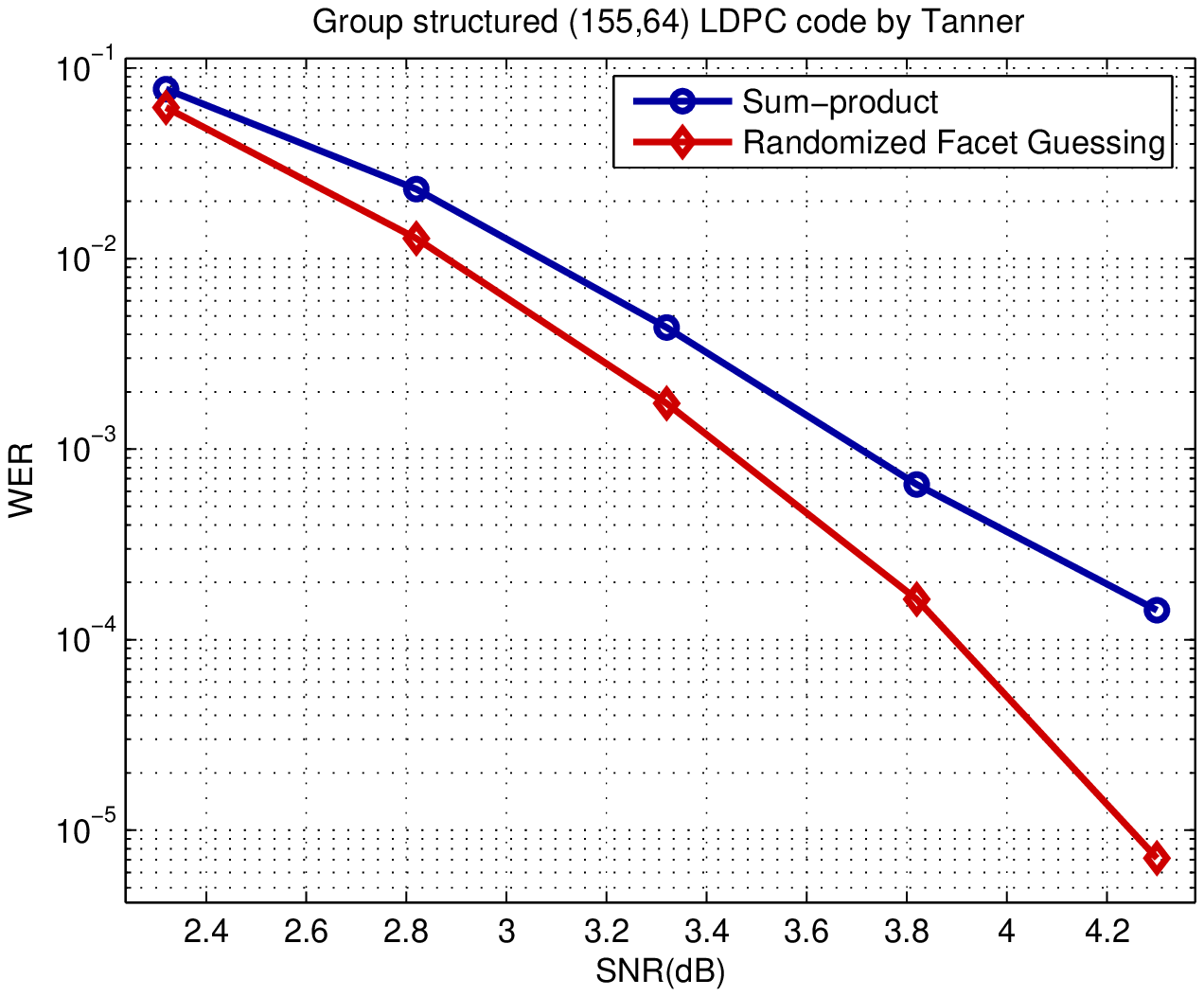} &
\widgraph{.38\textwidth}{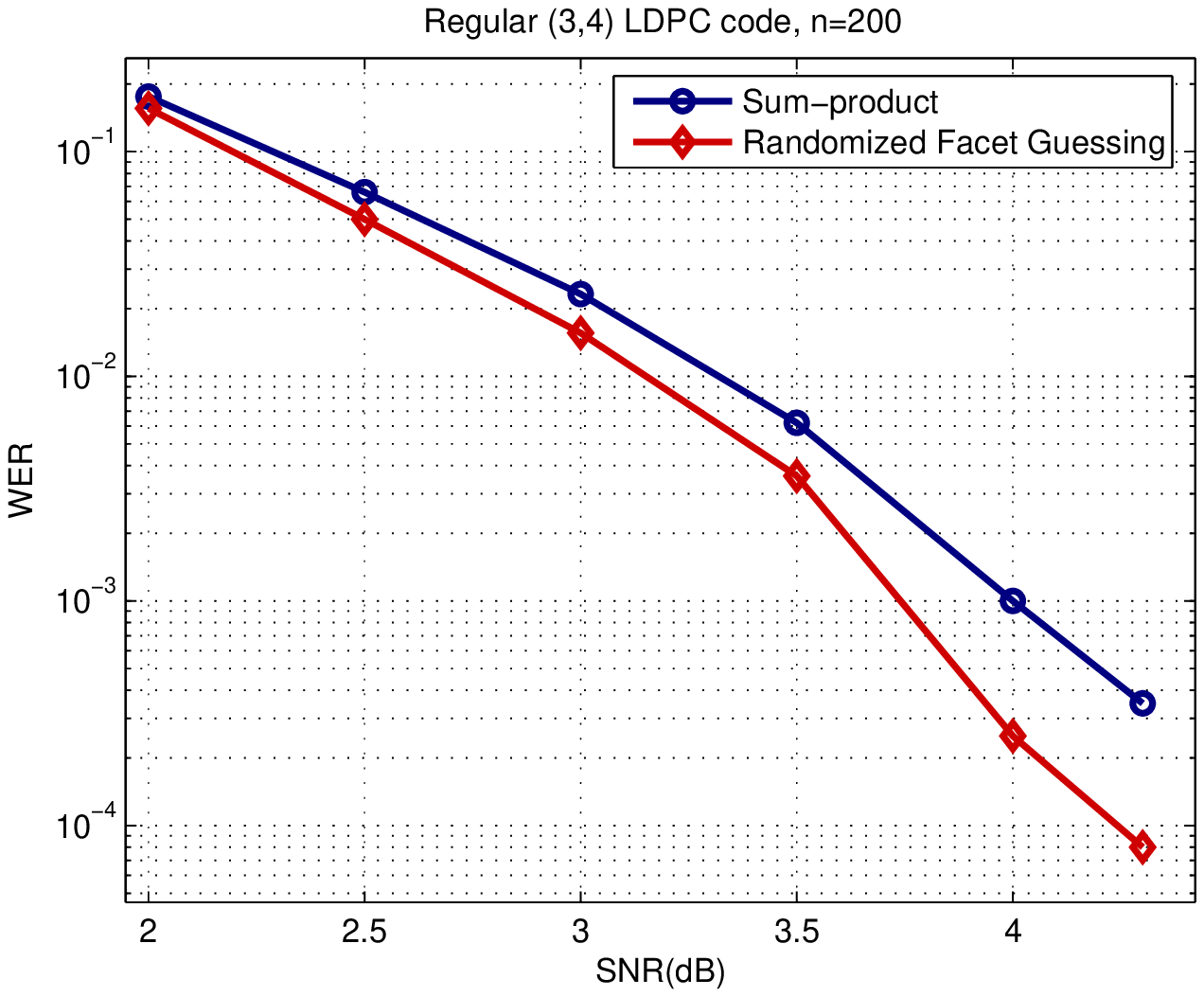} \\ (a) & (b)
\end{tabular}
\end{center}
\caption{Comparison of different decoding methods: standard
sum-product decoding, and randomized facet-guessing (RFG) with $N =
20$ iterations. The two panels show two different codes: (a) Tanner's
group-structured code.  (b) Random (3,4) LDPC code with $n=200$. }
\label{FigPerformance}
\end{figure*}

By using this characterization and Theorem~\ref{ThmPCWSize} for
expander codes, we obtain the following result:
\begin{corollary}
\label{CorExhaust} For expander codes, the EFG algorithm will 
always succeed if there are $\const_1$ fractional pseudocodewords with
likelihood higher than the ML codeword and $\const_1 <
\frac{\gamma_{cw}}{\gamma_{pc}}$.  Under this condition, each
iteration of RFG succeeds with constant probability \mbox{$p_{RFG}
\geq \frac{ \gamma_{cw} - \const_1 \gamma_{pc}} {2^{d_c-1}(1-R)+2}$.}
%will always succeed if there are less than or equal to
%\begin{eqnarray}
%\const_1 & \defn & \left \lfloor \frac{(1-\rate)d_c +1}{ \big(
%1-\rate - \alpha d_v/2 \big)d_c +d_v \alpha + (1 - \alpha)} \right
%\rfloor
%\end{eqnarray}
%fractional pseudocodewords with higher likelihood.
\end{corollary}
\mybeginproof
From Lemma~\ref{exhaustive}, the EFG algorithm fails if and only if
every facet in $|\actset_c|$ also contains another fractional
pseudocodeword with higher likelihood.  But for expander codes,
Lemma~\ref{LemUpperBound} yields that the size of the active
set of any fractional pseudocodeword is upper bounded as
\begin{eqnarray*}
|\actset_p| & \leq n \gamma_{pc}.
\end{eqnarray*}
while the size of active sets of any codeword is always
$|\actset_c|= n \gamma_{cw}$. Therefore, if there exist $\const_1$
fractional pseudocodewords with likelihood higher than $c$, the
total number of facets adjacent to these fractional
pseudocodewords is at most $\gamma_{pc} \const_1 n$. Therefore
when $ \gamma_{pc} \const_1 n < n \gamma_{cw}$ it is impossible to
completely cover $\actset_c$ and EFG succeeds. Also RFG at each
iteration selects a random facet and there are $(\gamma_{cw}
-\gamma_{pc}\const_1) n$ facets that contain $c$ but not any
fractional pseudocodeword with higher likelihood. The total number
of facets is $|\actset|= (2^{d_c-1}(1-R)+2) n $ and therefore each
iteration of RFG has probability of success larger than $\frac{
\gamma_{cw} - \const_1 \gamma_{pc}} {2^{d_c-1}(1-R)+2}$.
 \myendproof \\

Notice that this corollary only provides a worst case bound. Even
though there is a linear number of facets that contain the ML
codeword, we show that it will require a constant number of fractional
pseudocodewords to cover them.  This can only happen if the high
likelihood fractional pseudocodewords have their adjacent facets
non-overlapping and entirely contained in $A_c$.  More typically, one
could expect the facet guessing algorithm to work even if there are
many more fractional pseudocodewords with higher likelihoods.  Indeed,
our experimental results show that the RFG algorithm leads to a
significant performance gain for those codewords that are recovered
successfully by neither sum-product nor LP decoding.  As shown in
Figure~\ref{FigPerformance}, the gains are pronounced for higher SNR,
as high as $0.5$dB for the small blocklengths that we experimentally
tested.  The added complexity corresponds to solving a constant number
of LP optimizations; moreover, the extra complexity is required
\emph{only if} LP decoding fails.

%For expander codes, we have shown that $A_c$ will be much larger than
%$A_p$, and therefore, it is not necessary to exhaustively examine all
%the $|A_c \setminus A_p|$ possible facets. We will show that if one ML
%codeword has less than $c_2 = (1-\epsilon) c_1$ fractional
%pseudocodewords, a random selection of a facet from $|A_c \setminus
%A_p|$ will contain $A_c$ but not any $p \in \hat{p}$ with at least a
%constant probability.  (...randomized facet-guessing...)

%%%%%%%%%%%%%%%%%%%%%%%%%%%%%%%%%%%%%%%%%%%%%%%%%%%%%%%%%%%%%%

\section{Discussion}
We have investigated the structure of the polytope that underlies both
LP decoding and the sum-product algorithm.  We show that for expander
codes, every fractional pseudocodeword always has at least a constant
fraction of non-integral bits. We further proposed an decoding
algorithm, with complexity only a constant factor larger than that of
the LP decoder, and analyzed the performance gains that it achieves.
This theoretical analysis is supplemented with experimental results
showing gains for short to moderate block lengths, particularly at
high SNR.
%
%
%\texttt{TO BE WRITTEN}

%--------------------------------

%The postprocessing algorithm receives as input the relaxed linear program (described by the polytope in

%Consider the scenario where some codeword $c_1$ was transmitted over a noisy channel
%and the result of the LP relaxation
%We introduce the Random Active Facet postprocessing algorithm a

{\footnotesize{
\section*{Acknowledgment}
Work partially supported by NSF Grant DMS-0528488, and a UC-Micro
grant through Marvell Semiconductor.  }}


% http://www.ctan.org/tex-archive/macros/latex/contrib/supported/IEEEtran/bibtex
%\bibliographystyle{IEEEtran.bst}

%\bibliographystyle{mylatex8}
%{\footnotesize{

%\bibliography{BIBTEX_a_isit06}
%}}

{\footnotesize{

}}

\end{document}